\documentclass[11pt,twoside
]{article}

\usepackage{jae2016} 
\usepackage{graphicx}
\usepackage{subfigure}
\usepackage{psfrag}
\usepackage{amssymb}
\usepackage[spanish,activeacute,english]{babel}
\usepackage[latin1]{inputenc}
\usepackage[T1]{fontenc} 
\usepackage{ae,aecompl} 
\usepackage{latexsym}
\usepackage{verbatim}
\usepackage{amsmath}
\usepackage{stmaryrd}
\usepackage{amsfonts}
\usepackage{amssymb}
\usepackage{wasysym}
\usepackage[colorlinks=true,dvips]{hyperref}

\begin{document}
\myselectspanish
\vskip 1.0cm
\markboth{ D. Merlo \& S. Gurovich}%
{Nuevo algoritmo de identificacio\'n en SACAMAN}

\pagestyle{myheadings}

\title{Implementaci\'on de un nuevo algoritmo de identificaci\'on de fuentes estelares en el c\'odigo SACAMAN}

\author{
D. C. Merlo$^{1}$ \&
S. Gurovich$^{1,2}$
}

\affil{%
  (1) Observatorio Astron\'omico (UNC)\\
  (2) IATE (CONICET)\\
}

\begin{resumen}
Se presenta una versi\`on mejorada del m\`odulo de identificaci\`on de fuentes estelares en el c\`odigo cuasi-autom\`atico SACAMAN, el cual permite obtener magnitudes YZJHKs-VVV para un conjunto de objetos de inter\`es.
El procedimiento utiliza un algoritmo de proximidad, insumiendo un tiempo de procesamiento $\simeq$ 70  veces menor que el m\`etodo de aproximaciones y conteos sucesivos de la versi\`on anterior.
Este c\`odigo est\`a siendo utilizado en el estudio de variabilidad de estrellas de carbono pertenecientes al bulbo gal\`actico.
\end{resumen}

\begin{abstract} 
An improved version of the identification stellar sources module in the quasi-automatic SACAMAN code is presented, which allows to obtain YZJHKs-VVV magnitudes for a set of predefined objects.
The procedure uses a proximity algorithm $\simeq$ 70 less time-consuming than the previous method of successive and counting approximation.
This code is being applied to the study of variability of carbon stars belonging to the Galactic Bulge.
\end{abstract}

\section{Introducci\`on}
Las estrellas de carbono (CSs) son estrellas gigantes fr\`ias evolucionadas que presentan material circunestelar en forma de $shells$, granularidad amorfa, discos o nubes. Uno de sus fen\`omenos caracter\`isticos es la variabilidad y el an\`alisis de la misma permite explicar las propiedades f\`isicas y los procesos que tienen lugar en sus atm\`osferas, como as\`i tambi\`en determinar su estado evolutivo (Alksne et al. 1991).

El relevamiento VVV provee una excelente oportunidad para llevar adelante estudios precisos de variabilidad, ya que nos brinda fotometr\`ia infrarroja profunda y multi-\`epocas que permiten construir curvas de luz de alta calidad (Merlo 2015, 2016a,b). Debido a la gran cantidad de datos que el mismo brinda, elaboramos un c\`odigo  que permite obtener para tal fin las magnitudes YZJHKs de cada objeto de inter\`es.

\section{Procedimiento}

\begin{figure}[!ht]
  \centering
  \includegraphics[width=.30\textwidth]{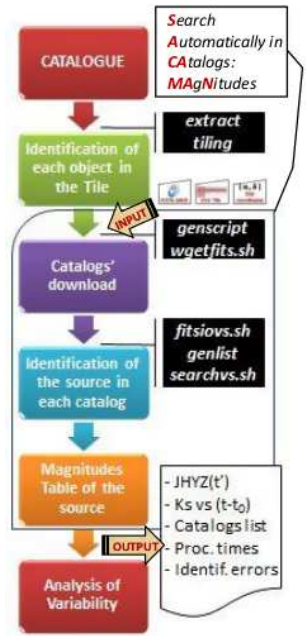}
  \caption{Diagrama de flujo del c\`odigo. En los rect\`angulos negros se indican los programas usados en cada etapa ($scripts.sh$ y c\`odigos $Fortran$). Ver detalles en Merlo (2016a).}
  \label{fig:ab1}
\end{figure}

En la Figura \ref{fig:ab1} se muestra el procedimiento general del c\`odigo. El mismo consta de tres m\`odulos: {\em Descarga $-$ Identificaci\`on $-$ Tabulaci\`on} (Merlo 2016a).

El procedimiento se inicia introduciendo en un $script$ gen\`erico el c\`odigo num\`erico que permite la descarga de los cat\`alogos, previamente gestionado en la plataforma CASU, el $tile$ de trabajo y las coordenadas de cat\`alogo de la fuente de inter\`es. Posteriormente se lo ejecuta en consola y autom\`aticamente lleva a cabo cada uno de los m\`odulos para lo cual requiere conectividad Internet permanente. Las corridas del c\`odigo fueron realizadas en grupos de 8 a 10 fuentes, debido a imprevisibles cortes de la conexi\`on a Internet.

\subsection{Cat\`alogos}

Los cat\`alogos fotom\`etricos utilizados fueron obtenidos a partir de las im\`agenes VVV ya procesadas puestas a disponibilidad del grupo colaborador por CASU ({\em Cambridge Astronomical Survey Unit}). Las mismas contienen las posiciones, los flujos y algunas mediciones de forma obtenidas con diferentes aperturas, que incluye la clasificaci\`on morfol\`ogica m\`as probable.

Cada unidad de observaci\`on que brinda el telescopio VISTA, de cuatro metros de di\`ametro, se denomina ``$tile$'', y consiste de 6 campos individuales de calado denominados ``$stacks$'' (ligeramente desplazados), cada uno de \`estos a su vez conformados por las 16 im\`agenes obtenidas en cada uno de los CCDs del instrumento, convenientemente superpuestos para producir la imagen final.

Toda la fotometr\`ia para este trabajo est\`a basada en los cat\`alogos $pawprint$ $stack$ CASU 1.3. Estos fueron elegidos en vez de los cat\`alogos $tiles$ m\`as profundos ya que, en estos \`ultimos, se han realizado correcciones fotom\`etricas a trav\`es del {\em CASU Vista Data Flow System} con el objetivo de aplanar diferencias espaciales que pudieran aparecer en cada uno de los detectores. Estas correcciones se han determinado para degradar aleatoriamente la precisi\`on fotom\`etrica de algunas fuentes puntuales (Irwin 2011).

La ventaja del uso de los cat\`alogos $stacks$ radica en el hecho de disponer de mayor cantidad (6x) de datos fotom\`etricos (\`epocas) para la fuente de inter\`es, lo cual redunda en una mejor calidad de las curvas de luz construidas a partir de ellas.

\subsection{Versiones}

Se han desarrollado una familia de versiones, las cuales se diferencian principalmente en el algoritmo de b\`usqueda de fuentes como as\`i tambi\`en en el tipo de archivos de cat\`alogo utilizado.

\begin{figure}[!ht]
  \centering
  \includegraphics[width=.35\textwidth]{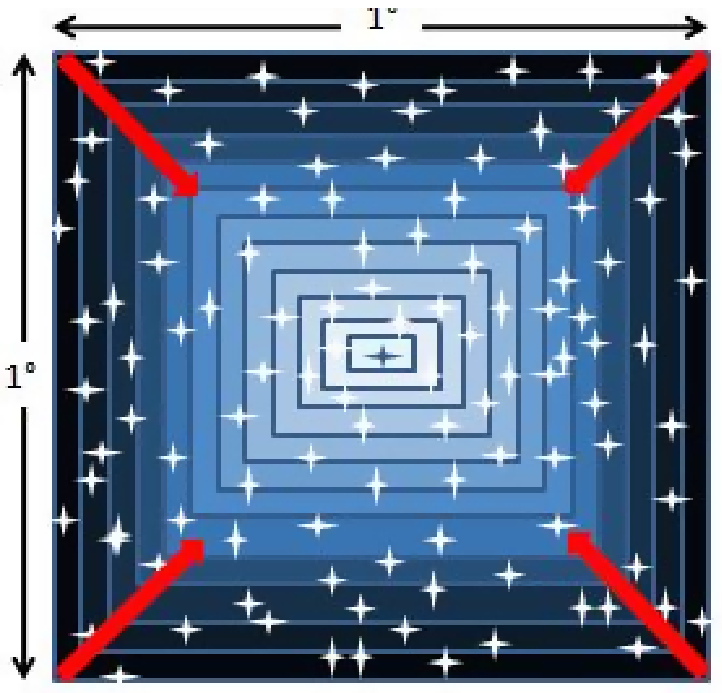}~\hfill
  \includegraphics[width=.33\textwidth]{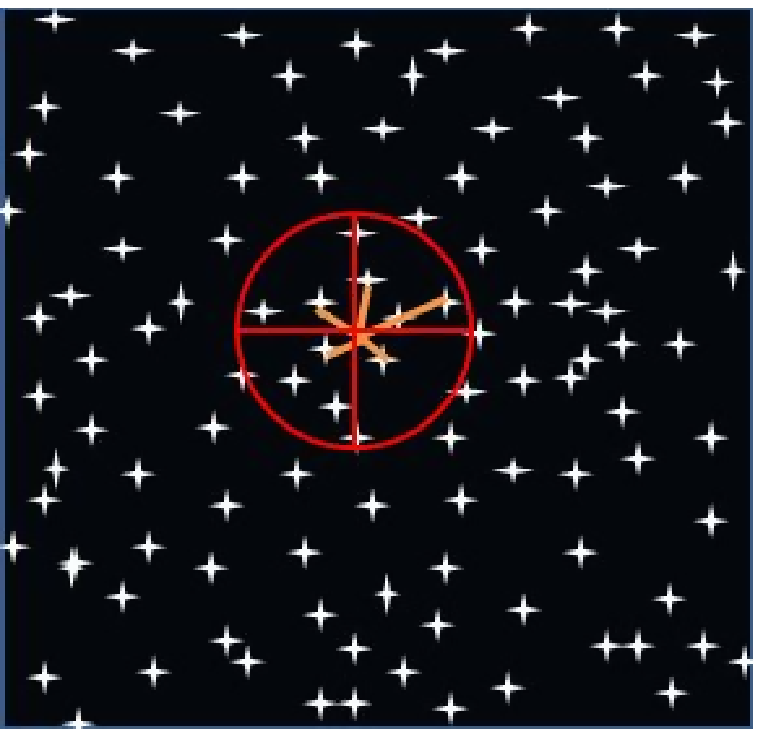}
  \caption{{\it a (Izquierda):} Descripci\`on esquem\`atica del procedimiento de aproximaciones y conteos sucesivos.
  \protect\\{\it b (Derecha):} Procedimiento actual de m\`inima distancia.}
  \label{fig:ab2}
\end{figure}

Las versiones 1.x utilizan el procedimiento de b\`usqueda por aproximaciones sucesivas, en el cual se va restringiendo el campo alrededor de la coordenada de cat\`alogo y contando las fuentes incluidas, deteni\`endose el proceso cuando no se encuentra ninguna fuente (ver Fig. \ref{fig:ab2}a).

En cambio las versiones 2.x hacen uso de un procedimiento de minimizaci\`on de distancia a las coordenadas de cat\`alogo. En el mismo se identifican las fuentes estelares halladas en un campo centrado en dichas coordenadas y se calculan las distancias angulares respectivas a ellas, seleccion\`andose aquella fuente que disponga del menor valor (ver Fig. \ref{fig:ab2}b).

A su vez, las versiones x.0 utilizan los cat\`alogos provenientes de las baldosas o $tiles$ (Merlo 2016a), mientras que las versiones x.5 los cat\`alogos $stacks$ (Merlo 2016b).

\section{Resultados}

En relaci\`on a los tiempos de procesamiento, y a modo de ejemplo, en un campo con 396 cat\`alogos $stacks$, el nuevo proceso de identificaci\`on (versi\`on 2.5) insumi\`o 7$^m$ ($\simeq$1 seg/cat), mientras que en otro campo, con 77 cat\`alogos $tiles$, el anterior proceso de identificaci\`on (versi\`on 1.0) necesit\`o 1,5$^h$ ($\simeq$70 seg/cat). 

\begin{figure}[!ht]
  \centering
  \includegraphics[width=.40\textwidth]{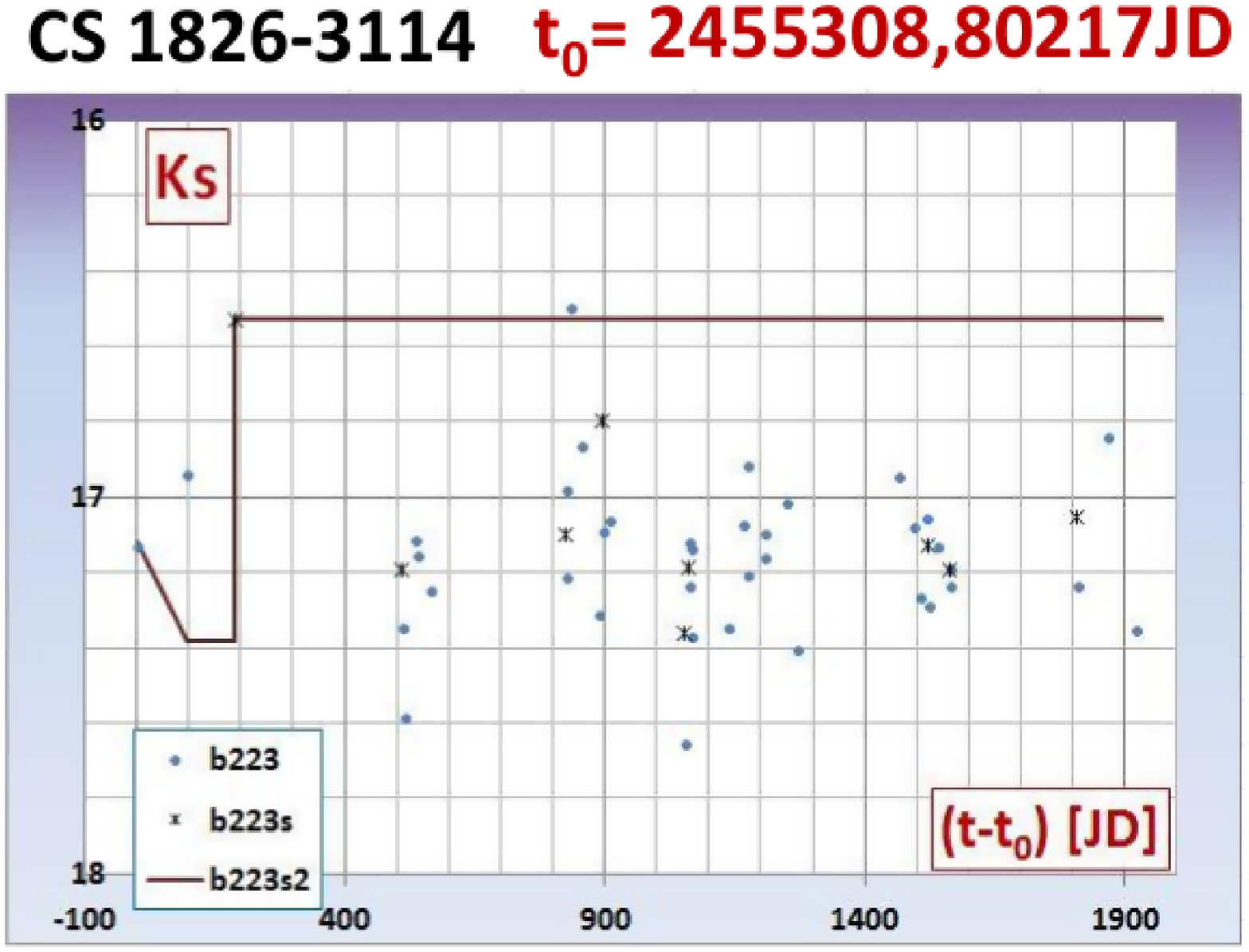}~\hfill
  \includegraphics[width=.40\textwidth]{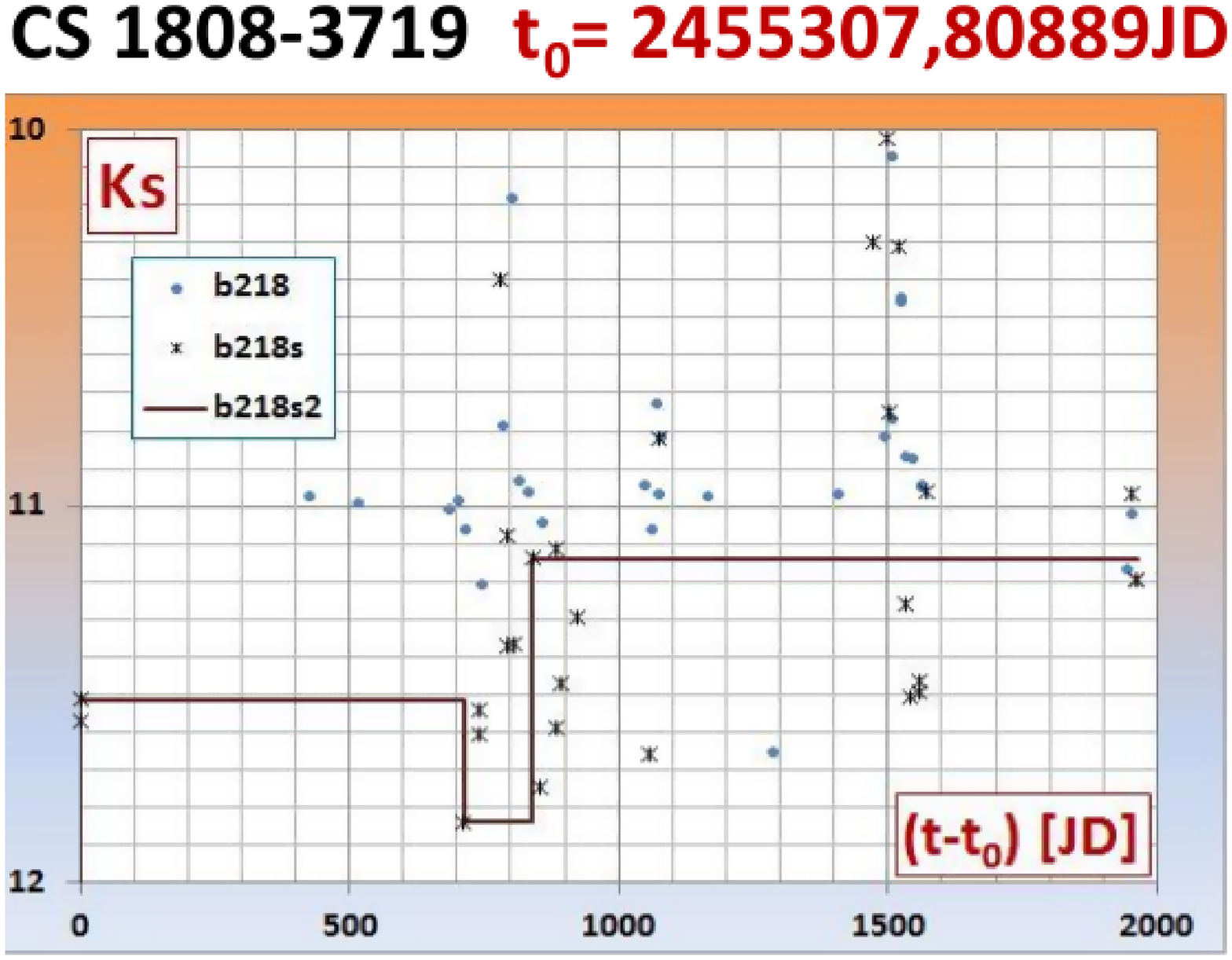}
  \caption{{\it a (Izquierda):} Variabilidad Ks de la estrella CS 1826-3114.
  \protect\\{\it b (Derecha):} Variabilidad Ks de la estrella CS 1808-3719.}
  \label{fig:ab3}
\end{figure}

Los gr\`aficos de la Fig. \ref{fig:ab3} muestran, en trazo continuo, los nuevos valores ({\bf s2}) de las magnitudes Ks de las CS halladas, comparados con las versiones anteriores 1.0 y 1.5 ({\bf s}). Las estrellas pertenecen a los $tiles$ del bulbo gal\`actico $b221$ y $b223$.

\section{Conclusiones}

Los tiempos de procesamiento obtenidos disminuyeron significativamente, logr\`andose adem\`as optimizar el proceso de detecci\`on e identificaci\`on eliminando muchos falsos positivos.

\section{Perspectivas futuras}

Este c\`odigo se encuentra en etapa de depuraci\`on y ampliaci\`on, las cuales una vez implementadas el mismo ser\`a puesto a disposici\`on de toda la comunidad interesada en utilizarlo.

Recientemente se ha finalizado el procedimiento de descarga y obtenci\`on de las magnitudes YZJHKs-VVV de todas las estrellas de carbono pertenecientes al bulbo de la V\`ia L\`actea, por lo que se ha iniciado el proceso de an\`alisis de variabilidad. Una vez finalizado, se llevar\`a adelante un pro\-ce\-di\-mien\-to similar con las CSs pertenecientes a la zona del disco gal\`actico cubierto por el relevamiento VVV.

Cabe destacar, finalmente, que este relevamiento tendr\`a pr\`oximamente una extensi\`on denominada VVV-X, el cual ampliar\`a un poco m\`as la zona cubierta por el VVV. Por ello tenemos previsto continuar nuestro estudio utilizando los nuevos y valiosos c\`atalogos que surgir\`an del mismo.

\begin{referencias}

\vskip .5cm

\reference Alksne, Z. et al., 1991, ``Properties of Galactic Carbon Stars'', Orbit Book Co.
\reference Alksnis, A. et al., 2001, Balt.A., 10, 1.
\reference Irwin, M.J., 2011, comunicaci\`on privada.
\reference Merlo, D., 2015, BAAA, 57, 111.
\reference Merlo, D., 2016a, 7VVV Workshop, 29-1/03, Antofagasta, Chile. \href{http://vvvsurvey.org/type/gallery/#jp-carousel-1101}{ Link}
\reference Merlo, D., 2016b, FoF2016, 29-1/04, C\`ordoba, Argentina. \href{http://fof.oac.uncor.edu/fof/posters/merlo.html}{ Link}
\reference Saito, R. et al., 2012, A\&A, 544, A147.
                                                                               
\end{referencias}

\end{document}